\documentclass{kapproc} 

\kluwerbib

\begin{document}

\articletitle{The QSO -- Ultraluminous Infrared Galaxies Connection}


\author{Sylvain Veilleux}
\affil{Department of Astronomy, University of Maryland}
\email{veilleux@astro.umd.edu}

\author{Dong-Chan Kim}
\affil{Institute of Astronomy and Astrophysics, 
Academia Sinica}
\email{kim@asiaa.sinica.edu.tw}

\author{David B. Sanders}
\affil{Institute for Astronomy, 
University of Hawaii}
\email{sanders@galileo.ifa.hawaii.edu}

\begin{abstract}
For the past several years, our group has pursued a vigorous
ground-based program aimed at understanding the nature of
ultraluminous infrared galaxies.  We recently published the results
from a optical/near-infrared spectroscopic survey of a large
statistically complete sample of ultraluminous infrared galaxies
(the "IRAS 1-Jy sample"). We now present the results from our
recently completed optical/near-infrared imaging survey of the 1-Jy
sample.  These data provide detailed morphological information on both
large scale (e.g., intensity and color profiles, intensity and size of
tidal tails and bridges, etc) and small scale (e.g., nuclear
separation, presence of bars, etc) that helps us constrain the initial
conditions necessary to produce galaxies with such high level of star
formation and/or AGN activity.  The nature of the interdependence
between some key spectroscopic and morphological parameters in our
objects (e.g., dominant energy source: super-starburst versus quasar,
nuclear separation, merger phase, star formation rate, and infrared
luminosity and color) is used to clarify the connection between
starbursts, ultraluminous infrared galaxies, and quasars.
\end{abstract}


\section{Introduction}

Ultraluminous infrared galaxies (ULIGs; log [L$_{\rm IR}$/L$_\odot$]
$\ge$ 12 by definition) may provide the clearest observational link
between galaxy mergers, starbursts and powerful AGN.  However, the
exact nature of ULIGs remains unclear. The most important questions,
and the ones we propose to address here, are: (1) {\em What is the
dominant energy source in ULIGs (starburst versus AGN)?}, and (2)
{\em Is there a evolutionary connection between ULIGs and quasars?}

In recent years, several surveys to faint flux levels in the IRAS
database have been carried out to search for luminous objects.
Arguably the most important of these studies is the `1~Jy' survey of
Kim (1995).  This study provides a complete list of the brightest
ULIGs with F[60~$\mu$m] $>$ 1~Jy which is not biased toward `warm'
quasar-like objects with large F[25 $\mu$m]/F[60 $\mu$m] ratios.  The
`1~Jy' sample contains 118 objects with $z$ = 0.02 -- 0.27 and
log~[L$_{\rm ir}$/L$_\odot$] = 12.00 -- 12.84.  The infrared
luminosities of these objects therefore truly overlap with the
bolometric luminosities of optical quasars.  Other surveys have
discovered objects of comparable luminosity at fainter flux levels as
well as a few `hyperluminous' objects at higher L$_{\rm ir}$. However,
the `1~Jy' sample contains the brightest objects at a given
luminosity, hence the best candidates for follow-up studies. 

The results of our analysis of the IRAS database on the `1~Jy' sample
were published in Kim \& Sanders (1998) and are discussed in these
proceedings (Sanders contribution). The present paper summarizes the
results from our ground-based follow-up surveys. In Section 2, we
discuss the results from our optical and near-infrared spectroscopy of
this sample, and in Section 3, the preliminary results from our
optical and near-infrared imaging survey. Section 4 summarizes our
conclusions.

\section{Optical and Infrared Spectroscopy of the 1 Jy Sample}

The results from our optical and near-infrared spectroscopic surveys
were published recently in Veilleux, Kim, \& Sanders (1999a) and
Veilleux, Sanders, \& Kim (1999b), respectively.  The main conclusions
of this analysis are the following:

\begin{itemize}

\item[1.]  The fraction of luminous infrared galaxies with Seyfert
characteristics increases rapidly with increasing $L_{\rm ir}$.  About
30\% of the ULIGs host a Seyfert 1 or Seyfert 2 nucleus.  For $L_{\rm
ir} > 10^{12.3}\ L_\odot$, this fraction is nearly 50\%.

\item[2.]  From 50\% to 70\% of the Seyfert 2s in our sample show
signs of an AGN in the near-infrared (e.g., broad-line region or
strong [Si~VI] 1.962 $\mu$m), therefore confirming the detection of
genuine AGN in these objects. In contrast, none of the optically
classified LINERs and H~II galaxies in our near-infrared sample shows
any obvious signs of an energetically important AGN.

\item[3.] Combining our optical and near-infrared results, we find that the
fraction of objects with bonafide AGN is 25-30\% for the 1 Jy sample
of ULIGs and 35-50\% for objects with $L_{\rm ir} > 10^{12.3}\ L_\odot$.

\item[4.] Comparisons of the dereddened emission-line
luminosities of the optical or obscured BLRs detected in the ULIGs of
the 1-Jy sample with those of optical quasars indicate that the
obscured AGN/quasar in ULIGs is the main source of energy in at
least 15 -- 25\% of all ULIGs in the 1-Jy sample. This fraction is
closer to 30 -- 50\% among ULIGs with $L_{\rm ir} > 10^{12.3}\
L_\odot$. 

\end{itemize}

These results are compatible with those from recent mid-infrared
spectroscopic surveys carried out with {\it ISO} (e.g., Genzel et
al. 1998). Indeed, a detailed object-by-object comparison of the
optical and mid-infrared classifications shows an excellent agreement
between the two classification methods (Lutz, Veilleux, \& Genzel
1999).  These results suggest that strong nuclear activity, once
triggered, quickly breaks the obscuring screen at least in certain
directions, thus becoming detectable over a wide wavelength range.

\section{Optical and Infrared Imaging of the 1 Jy Sample}

We now have high signal-to-noise ratio, sub-arcsecond resolution R and
K$^\prime$ images of all `1~Jy' sources.  Additionnal spectra were
obtained of several sources in the field to identify them (stars
versus galaxies) and determine if they are involved in the ULIG
event. Our preliminary analysis of these data suggests the following
tantalizing trends:

\begin{itemize}

\item[1.] As found in previous studies, the large majority ($>$
95\%) of the optical and near-infrared images show signs of a strong
tidal interaction/merger in the form of distorted or double nuclei,
tidal tails, bridges, and overlapping disks.

\item[2.]  The small mean nuclear separation ($<$ 3 kpc) of the ULIGs
examined so far suggests that the majority of these galaxies are in a
terminal stage of a merger. These mergers generally involve {\em two}
large (0.5 -- 2 L$^*$) galaxies. Multiple mergers are seen in only 4
of the 118 systems.

\item[3.] These galaxies span a broad range of total (= nuclear +
host) luminosities. In advanced mergers, L$_{\rm R}$(tot) $\sim$ 1.5
L$^*$ and L$_{{\rm K}^\prime}$(tot) $\sim$ 3 L$^*$ on average but with
a lot of scatter.

\item[4.] Roughly 30\%/10\%/60\% of the R-band surface brightness
profiles are well fitted by a elliptical-like R$^{1/4}$-law /
exponential disk / neither or both. The percentage of poor fits is
likely to decrease when we examine our data at K$^\prime$, where 
the effects of dust obscuration and star formation are less important. 

\item[5.] Using a classification scheme first proposed by Surace
(1998) and based on the results of published numerical simulations, we
classify all our objects according to morphology: I. Pre-contact --
relatively unperturbed and separate disk. II. First contact --
overlapping disks but no evidence for strong bars or tidal
tails. III. Pre-merger -- two distinct galaxies with well-developed
tidal tails and brigdes. (a) apparent separation $>$ 10 kpc and (b)
apparent separation $<$ 10 kpc. IV. Merger -- single nucleus with
prominent tidal tails. (a) diffuse nucleus with L$_{{\rm K}^\prime}$(2
kpc)/L$_{{\rm K}^\prime}$(tot) $<$ 1/3. (b) compact nucleus with
L$_{{\rm K}^\prime}$(2 kpc)/L$_{{\rm K}^\prime}$(tot) $>$ 1/3. V. Old
merger -- no obvious signs of tidal tails but disturbed central
morphology.  We find that all Seyfert 1s and most of the Seyfert 2s
are advanced mergers either based on their morphology (classes IVb or
V; Fig. 1) or their nuclear separation ($<$ 5 kpc, generally;
Fig. 2). A similar result is found when we consider the `warm' objects
with $f_{20}/f_{60}$ $>$ 0.2 (Fig. 3).

\end{itemize}

\begin{figure}[ht]
\vskip 3in
\caption{Optical spectral type versus morphological class. See text for
a description of the morphological classification.}
\end{figure}

\begin{figure}[ht]
\vskip 3in
\caption{Optical spectral type versus nuclear separation.}
\end{figure}

\begin{figure}[ht]
\vskip 3in
\caption{IRAS $f_{20}/f_{60}$ colors versus morphological classification.}
\end{figure}

\section{Summary}

The results from our spectroscopic survey of the 1 Jy sample of ULIGs
indicate that the fraction of ULIGs powered by quasars increases with
increasing infrared luminosity, reaching a value of 30 -- 50 \% for
$L_{\rm ir} > 10^{12.3}\ L_\odot$. The preliminary results from our
imaging survey of the same sample suggest trends between merger phase,
infrared colors, and the presence of an AGN. Objects with `warm'
quasar-like infrared colors show signs of AGN activity and are
generally found in advanced mergers, based not only on the apparent
nuclear separation but also on the morphology of the tidal
tails. These results suggest that the evolutionary sequence `cool'
ULIGs $\rightarrow$ `warm' ULIGs $\rightarrow$ quasars applies to at
least some (though admittedly probably not all) ULIGs and quasars.

\begin{acknowledgments}
S. V. gratefully acknowledges the financial support of NASA through
LTSA grant NAG 56547.
\end{acknowledgments}

\clearpage


\begin{chapthebibliography}{1}

\bibitem[Genzel et al. 1998]{gen98}
Genzel, R. et al. 1998, {\it Ap. J.}, {\bf 498}, 579
\bibitem[Kim \& Sanders 1998]{k98}
Kim, D.-C., \& Sanders, D. B. 1998, {\it Ap. J. Suppl.}, {\bf 119}, 41
\bibitem[Lutz et al. 1999]{l98}
Lutz, D., Veilleux, S., \& Genzel, R. 1998, {\it Ap. J. (Letters)}, {\bf  517}, L13
\bibitem[Sanders \& Mirabel 1996]{s96}
Sanders, D. B., \& Mirabel, I. F. 1996, AR\&A, {\bf 34}, 725
\bibitem[Sanders et al. 1988]{s88}
Sanders, D. B., et al. 1988, {\it Ap. J.}, {\bf 325}, 74
\bibitem[Veilleux et al. 1999a]{v99a}
Veilleux, S., Kim, D.-C., \& Sanders, D. B. 1999a, {\it Ap. J.}, {\bf 522}, 113
\bibitem[Veilleux et al. 1999b]{v99b}
Veilleux, S., Sanders, D. B., \& Kim, D.-C. 1999b, {\it Ap.~J.}, {\bf 522}, 139
\end{chapthebibliography}
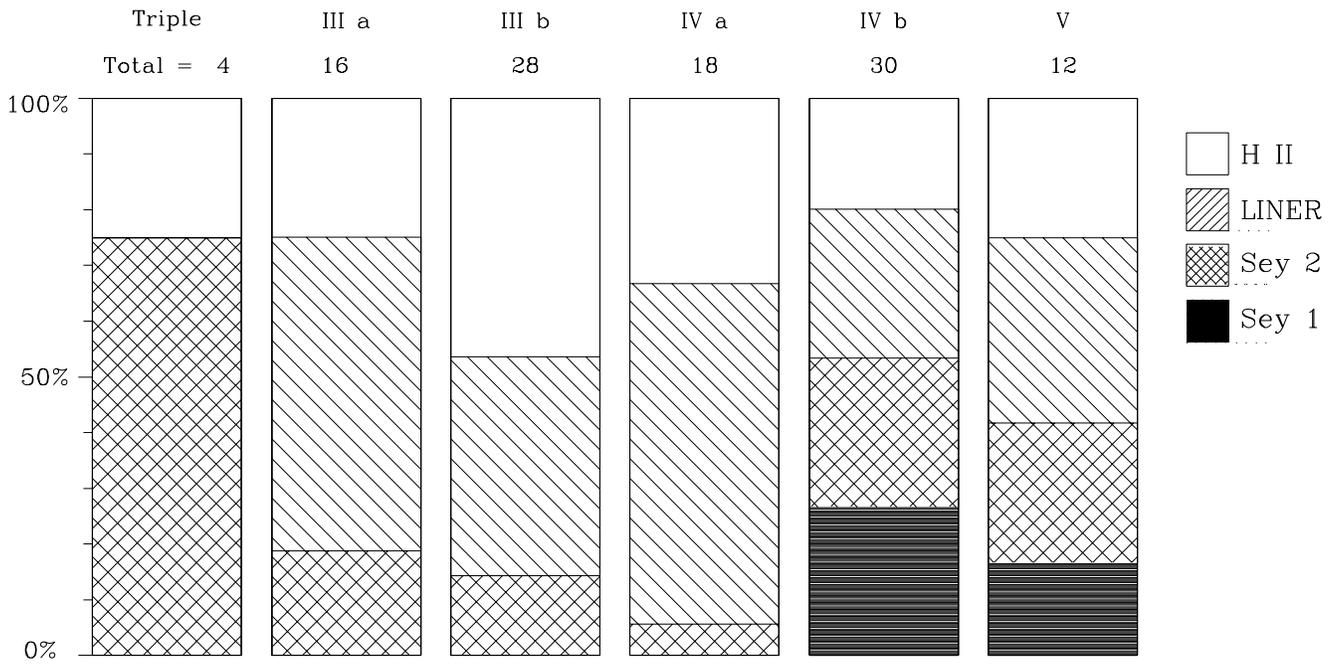
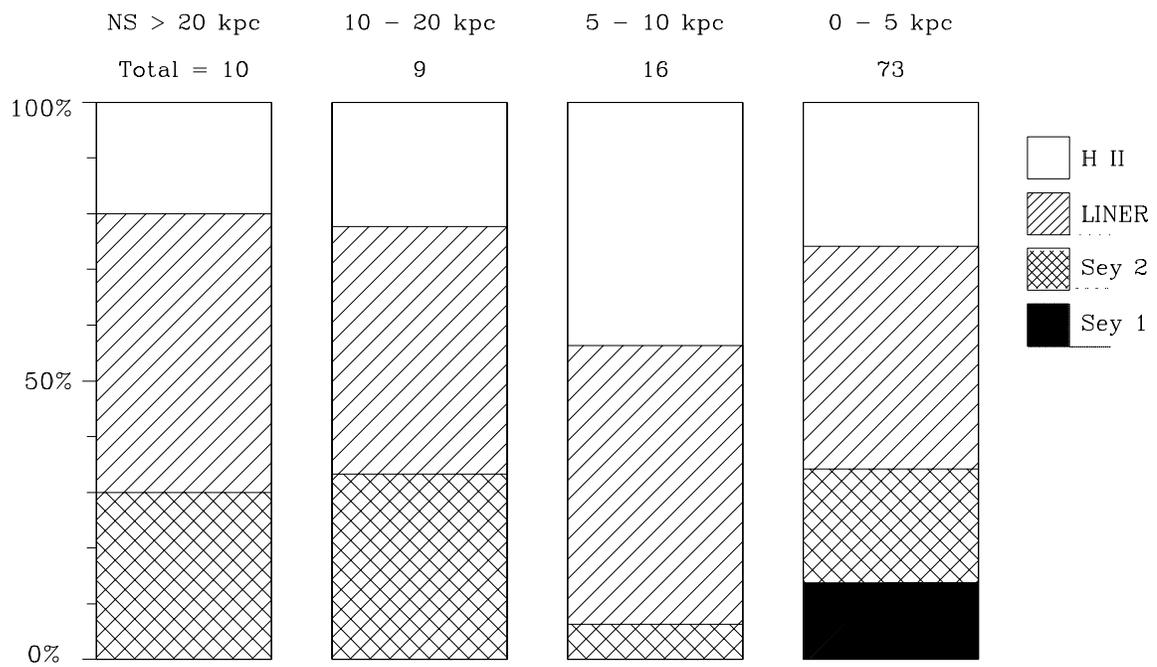
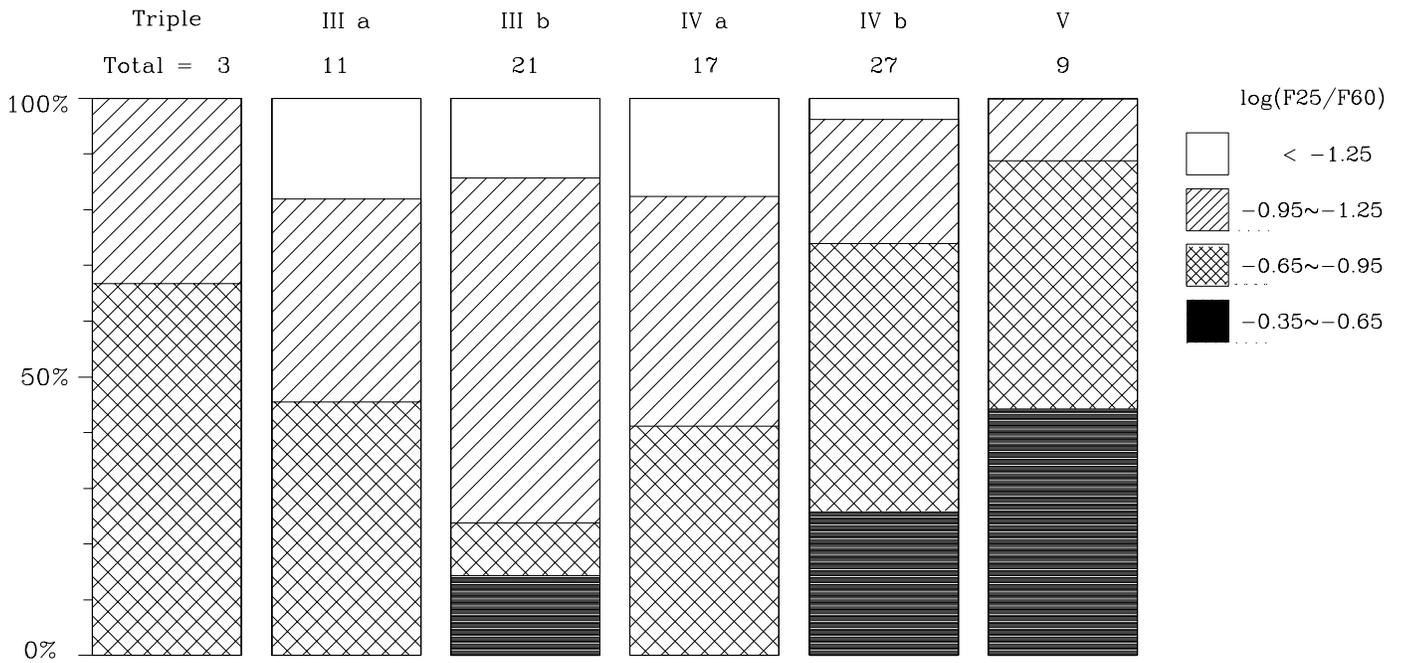
\end{document}